\documentclass[aps,prl,preprint,tightenlines,superscriptaddress,showpacs,byrevtex]{revtex4}

\usepackage{graphicx}
\usepackage{dcolumn}
\usepackage{bm}
\usepackage{epsfig}

\graphicspath{{ps}}

\begin{document}

\preprint{\vbox{
\hbox{\hfil BELLE-CONF-0419}
\hbox{\hfil ICHEP04 11-0665}
}}



\title{\quad\\[0.5cm] Measurement of Polarization and 
Triple-Product Correlations in $B \to \phi K^{*}$ Decays}
       


\affiliation{Aomori University, Aomori}
\affiliation{Budker Institute of Nuclear Physics, Novosibirsk}
\affiliation{Chiba University, Chiba}
\affiliation{Chonnam National University, Kwangju}
\affiliation{Chuo University, Tokyo}
\affiliation{University of Cincinnati, Cincinnati, Ohio 45221}
\affiliation{University of Frankfurt, Frankfurt}
\affiliation{Gyeongsang National University, Chinju}
\affiliation{University of Hawaii, Honolulu, Hawaii 96822}
\affiliation{High Energy Accelerator Research Organization (KEK), Tsukuba}
\affiliation{Hiroshima Institute of Technology, Hiroshima}
\affiliation{Institute of High Energy Physics, Chinese Academy of Sciences, Beijing}
\affiliation{Institute of High Energy Physics, Vienna}
\affiliation{Institute for Theoretical and Experimental Physics, Moscow}
\affiliation{J. Stefan Institute, Ljubljana}
\affiliation{Kanagawa University, Yokohama}
\affiliation{Korea University, Seoul}
\affiliation{Kyoto University, Kyoto}
\affiliation{Kyungpook National University, Taegu}
\affiliation{Swiss Federal Institute of Technology of Lausanne, EPFL, Lausanne}
\affiliation{University of Ljubljana, Ljubljana}
\affiliation{University of Maribor, Maribor}
\affiliation{University of Melbourne, Victoria}
\affiliation{Nagoya University, Nagoya}
\affiliation{Nara Women's University, Nara}
\affiliation{National Central University, Chung-li}
\affiliation{National Kaohsiung Normal University, Kaohsiung}
\affiliation{National United University, Miao Li}
\affiliation{Department of Physics, National Taiwan University, Taipei}
\affiliation{H. Niewodniczanski Institute of Nuclear Physics, Krakow}
\affiliation{Nihon Dental College, Niigata}
\affiliation{Niigata University, Niigata}
\affiliation{Osaka City University, Osaka}
\affiliation{Osaka University, Osaka}
\affiliation{Panjab University, Chandigarh}
\affiliation{Peking University, Beijing}
\affiliation{Princeton University, Princeton, New Jersey 08545}
\affiliation{RIKEN BNL Research Center, Upton, New York 11973}
\affiliation{Saga University, Saga}
\affiliation{University of Science and Technology of China, Hefei}
\affiliation{Seoul National University, Seoul}
\affiliation{Sungkyunkwan University, Suwon}
\affiliation{University of Sydney, Sydney NSW}
\affiliation{Tata Institute of Fundamental Research, Bombay}
\affiliation{Toho University, Funabashi}
\affiliation{Tohoku Gakuin University, Tagajo}
\affiliation{Tohoku University, Sendai}
\affiliation{Department of Physics, University of Tokyo, Tokyo}
\affiliation{Tokyo Institute of Technology, Tokyo}
\affiliation{Tokyo Metropolitan University, Tokyo}
\affiliation{Tokyo University of Agriculture and Technology, Tokyo}
\affiliation{Toyama National College of Maritime Technology, Toyama}
\affiliation{University of Tsukuba, Tsukuba}
\affiliation{Utkal University, Bhubaneswer}
\affiliation{Virginia Polytechnic Institute and State University, Blacksburg, Virginia 24061}
\affiliation{Yonsei University, Seoul}
  \author{K.~Abe}\affiliation{High Energy Accelerator Research Organization (KEK), Tsukuba} 
  \author{K.~Abe}\affiliation{Tohoku Gakuin University, Tagajo} 
  \author{N.~Abe}\affiliation{Tokyo Institute of Technology, Tokyo} 
  \author{I.~Adachi}\affiliation{High Energy Accelerator Research Organization (KEK), Tsukuba} 
  \author{H.~Aihara}\affiliation{Department of Physics, University of Tokyo, Tokyo} 
  \author{M.~Akatsu}\affiliation{Nagoya University, Nagoya} 
  \author{Y.~Asano}\affiliation{University of Tsukuba, Tsukuba} 
  \author{T.~Aso}\affiliation{Toyama National College of Maritime Technology, Toyama} 
  \author{V.~Aulchenko}\affiliation{Budker Institute of Nuclear Physics, Novosibirsk} 
  \author{T.~Aushev}\affiliation{Institute for Theoretical and Experimental Physics, Moscow} 
% ...
\collaboration{The Belle Collaboration}

\begin{abstract}
We present a measurement of the decay amplitudes and 
triple-product correlations in $B \to \phi K^*$ decays based on 
140~fb$^{-1}$ of data recorded at the $\Upsilon(4S)$ resonance with
the Belle detector at the KEKB $e^{+} e^{-}$ storage ring.
The decay amplitudes for the different $\phi K^{*}$ helicity states  
are measured from the angular distributions of final state particles 
in the transversity basis. The longitudinal and transverse complex amplitude moduli
and angles are $|A_0|^2 = 0.51 \pm 0.06 \pm 0.04$, 
$|A_\perp|^2 = 0.24 \pm 0.06 \pm 0.03$, 
$\arg(A_\parallel) = -2.21 \pm 0.22 \pm 0.05$~rad, and
$\arg(A_\perp) = 0.72 \pm 0.21 \pm 0.06$~rad. 
The $T$-violating asymmetries through triple-product correlations 
are measured to be consistent with zero.
\end{abstract}

\maketitle

\tighten

{\renewcommand{\thefootnote}{\fnsymbol{footnote}}}
\setcounter{footnote}{0}






The decay $B \to \phi K^*$ involves the FCNC $b\to s$ transition that is forbidden to first order 
in the Standard Model (SM) but can proceed by second order penguin and box diagrams.
This process therefore
provides information on the Cabibbo-Kobayashi-Maskawa matrix element $V_{ts}$ \cite{ref:ckm}
and is sensitive to physics beyond the SM.
The decay $B \to \phi K^*$ is a mixture of $CP$-even and $CP$-odd states, and
polarization measurements allow us to project out the different $CP$ states statistically.
Measurements of the $T$-odd triple-product correlations provide information
about $T$-violating asymmetries \cite{ref:TP-datta} as well as observables that are 
sensitive to New Physics (NP) \cite{ref:NP-bounds}. 
Our previous measurement \cite{ref:phik_prl} and a recent report by {\sc BaBar} \cite{ref:babar_ex0406063}
both suggest that the fraction of longitudinal polarization component differs from the naive SM prediction.
This provides a strong motivation to update our study with a larger dataset.

In this paper, we present a measurement of the decay amplitudes in 
$B^+ \to \phi K^{*+}$ and  $B^{0} \to \phi K^{*0}$ decays by a full three-dimensional angular analysis.
The charge conjugate modes are included everywhere unless otherwise specified.
We also report the measurements of triple-product correlations and related $T$-violating asymmetries,
as well as the observables that are sensitive to NP.


This analysis is based on a data set with an integrated luminosity of
140~fb$^{-1}$ taken at the $\Upsilon(4S)$ resonance
recorded by the Belle detector \cite{ref:Belle}
at the KEKB  $e^{+}e^{-}$ collider \cite{ref:KEKB}.
This luminosity corresponds to $152 \times 10^{6}$  
produced $B\bar{B}$ pairs. The beam energies are 8 GeV for $e^-$ and 3.5 GeV for $e^+$.

The Belle detector is a general purpose magnetic 
spectrometer equipped with a 1.5~T superconducting solenoid magnet. 
Charged tracks are reconstructed in a Central Drift Chamber (CDC)
and a Silicon Vertex Detector (SVD). 
Photons and electrons are identified using a CsI(Tl) Electromagnetic Calorimeter (ECL) 
located inside the magnet coil. 
Charged particles are identified using specific ionization ($dE/dx$)
measurements in the CDC as well as information from Aerogel Cherenkov 
Counters (ACC) and Time of Flight Counters (TOF). 
A kaon likelihood ratio, $P(K/\pi) = \mathcal{L}_K /(\mathcal{L}_K + \mathcal{L}_\pi)$, has values 
between 0 (likely to be a pion) and 1 (likely to be a kaon), where 
$\mathcal{L}_{K(\pi)}$ is derived from the $dE/dx$, ACC and TOF measurements.


Candidate $\phi \to K^+K^-$ decays are found by selecting pairs of 
oppositely charged tracks that are not pion-like ($P(K/\pi)>0.1$).
The vertex of the candidate charged tracks is required to be consistent with 
the interaction point (IP) to remove poorly measured tracks.
In addition, candidates are required to have a $K^+K^-$
invariant mass that is less than 10 MeV/$c^2$ from the nominal $\phi$ meson mass. 

Pairs of oppositely charged tracks are used
to reconstruct $K^0_S \to \pi^+\pi^-$ decays.
The $\pi^+\pi^-$ vertex is required to be displaced from 
the IP by a minimum transverse distance of 0.22~cm for
high momentum ($>1.5$ GeV/$c$) candidates and 0.08~cm for
those with momentum less than 1.5~GeV/$c$.
The direction of the pion pair momentum must agree with
the direction defined by the IP and the vertex displacement within 0.03 rad for 
high-momentum candidates, and within 0.1 rad for the remaining candidates.

Charged tracks with $P(K/\pi)>0.4$ ($<0.9$) are considered to be kaons (pions).
For $\pi^{0} \to \gamma\gamma$, a minimum photon energy of 50 MeV
is required and the $\gamma\gamma$ invariant mass must be less than 16 MeV/$c^2$
from the nominal $\pi^{0}$ mass. 
$K^{*}$ candidates are reconstructed in three decay modes: 
$K^{* 0} \to K^{+} \pi^{-}$, 
$K^{* +} \to K^{+} \pi^{0}$ 
and $K^{* +} \to K^{0}_{S} \pi^{+}$.
The invariant mass of the $K^{*}$ candidate is required to be less than 
70 MeV/$c^2$ from the nominal $K^{*}$ mass.

A $B$ meson is reconstructed from a pair of $\phi$ and 
$K^*$ candidates and identified by the
 energy difference 
$\Delta E \equiv E_B^{\rm cms} - E_{\rm beam}^{\rm cms}$, 
and the beam constrained mass $M_{\rm bc} \equiv \sqrt{(E^{\rm cms}_{\rm beam})^2 - (p_B^{\rm cms})^2}$.
$E_{\rm beam}^{\rm cms}$ is the beam energy in the center-of-mass system (cms) of the $\Upsilon(4S)$ resonance,
and $E_B^{\rm cms}$ and $p_B^{\rm cms}$ are the cms energy and momentum 
of the reconstructed $B$ candidate.
The $B$-meson signal window is defined as 5.27 GeV/$c^2$ $< M_{\rm bc} <$ 5.29 GeV/$c^2$
and $|\Delta E| <$ 45 MeV.
The signal window is enlarged to $-100$ MeV $<\Delta E<$ $80$ MeV for 
$B^+ \to \phi K^{*+} (K^{*+} \to K^{+} \pi^{0})$ 
because of the impact of shower leakage on $\Delta E$ resolution.
An additional requirement $\cos\theta_{K^*}<0.8$
is applied to reduce low momentum $\pi^0$ background,
where $\theta_{K^*}$ is the angle between the $K^*$ direction and 
its daughter kaon defined in the $K^*$ rest frame.
In the signal window, about 1\% of the events have multiple candidates.
We choose the best candidate according to the
goodness of fit of the $B$ meson's vertex.


The dominant background is continuum 
$e^+e^- \to q\overline{q}$ production ($q=u,d,c,s$). 
Several variables 
are used to exploit the differences between 
the event shapes for continuum $q\overline{q}$ production
(jet-like) and for $B$ decay (spherical) in the cms frame of 
the $\Upsilon(4S)$ \cite{ref:continuum_suppression}.
These variables are combined into a single likelihood ratio 
$R_s = {\cal L}_s/({\cal L}_s + {\cal L}_{q\overline{q}})$, where 
${\cal L}_s$ (${\cal L}_{q\overline{q}}$) denotes the signal (continuum)
likelihood. The requirements on the likelihood ratio are determined 
by a figure of merit study that includes a dependence on the $B$-tagging quality \cite{ref:tagging}. 

Backgrounds from other $B$ decay modes such as $B \to KKK^{*}$, 
$B \to f_0(980) K^{*}(f_0\to K^+K^-)$, $B \to \phi K\pi$, $B \to KKK\pi$, and 
feed-across between $\phi K^{*}$ and $\phi K$ decay channels are studied. 
The contributions from $B \to KKK^{*}$ and $B \to f_0(980) K^{*}(f_0\to K^+K^-)$ 
are estimated from the $K^+K^-$ invariant mass distribution.
The $K^+K^-$ mass distribution for $B \to KKK^{*}$ is determined by Monte Carlo (MC) simulation
assuming three-body phase space decay. 
The shape for $f_0(980)$ is obtained from MC, where a $S$-wave Breit-Wigner with a
40 MeV/$c^2$ intrinsic width is assumed \cite{ref:PDG}.
These contributions are estimated separately by fits to the events outside of the $\phi$ mass region.
The contribution from $B \to KKK^{*}$ is estimated to be $5-10$\% \cite{ref:range} of the signal yield
depending on $K^*$ decay modes.
The $B \to f_0 K^{*}$ contribution is estimated to be $4-11$\%.
Contamination from four-body $B \to KKK\pi$ decays is checked by performing 
fits to the non-resonant region of $K^+K^-$ and $K\pi$ mass.
It is found to be very small and is neglected. 
To remove the contamination of $\phi K$ decays, they are explicitly
reconstructed and removed from $\phi K^*$ candidates.


The signal yields ($N_s$) are extracted by 
extended unbinned maximum-likelihood fits 
performed simultaneously in $\Delta E$ and $M_{bc}$.
The signal probability density
functions (PDFs) are products of Gaussians in $\Delta E$ and $M_{\rm bc}$. 
The means and widths are 
verified using $B \to J/\psi K^*$ decays.
Additional bifurcated Gaussians (Gaussians with different widths 
on either side of the mean) are used to model 
the tails in the $\Delta E$ distribution.

\begin{figure}[!htb]
\begin{center}
\resizebox*{3in}{2.5in}{\includegraphics{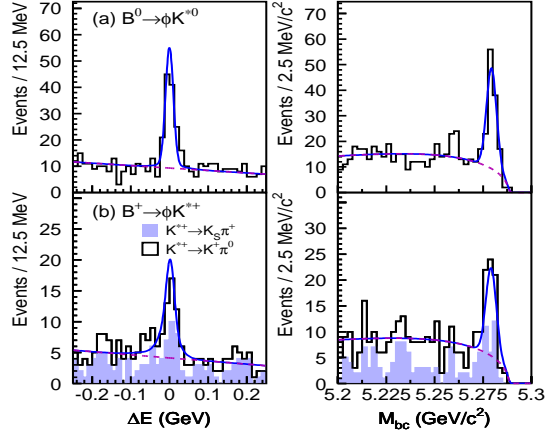}}
\end{center}
\caption{ Distributions of $\Delta E$ ($M_{\rm bc}$) with fit results for
the events in the $M_{bc}$ ($\Delta E$) signal window. 
The continuum background component is shown by the dashed curves. }
\label{fig:demb-projections}
\end{figure}

The PDF shape for the continuum events are parameterized by an ARGUS function in
$M_{\rm bc}$ and a linear function in $\Delta E$. The parameters of the functions are 
determined by a fit to the events in the signal sideband.  The sideband region
is defined by $M_{\rm bc} < 5.265$~GeV/c$^2$
and $|\Delta E| > 0.08$~GeV ($\Delta E > 0.10$~GeV and $\Delta E < -0.12$~GeV) 
for the modes without (with) a $\pi^0$ in the final state, respectively.
The signal and background yields are allowed to float in the fit while other PDF 
parameters are fixed.
The measured signal yields and purities are summarized in Table~\ref{tab:yields}.
The distributions of $\Delta E$ and $M_{\rm bc}$ are shown in Fig.~\ref{fig:demb-projections}.


\begin{table}[!htb]
\caption{Number of events observed in the signal box ($N_{ev}$), 
signal yields ($N_s$) obtained by fits after $KKK^*$ background subtraction, 
and signal purities.}
\label{tab:yields}	 
\begin{tabular}{lccc}
\hline 
\hline
Mode 			        & $N_{ev}$ & $N_{s}$ 		  & Purity \\
\hline
$B^{0} \to \phi K^{*0}$         & 167      & $97^{+12}_{-11}$     & $58.1^{+7.2}_{-6.6}$\% \\
\hline
$B^{+} \to \phi K^{*+} (K^{*+}\to K^{0}_{S}\pi^{+})$     & 40       & $20.4^{+5.9}_{-5.2}$ & $51^{+15}_{-13}$\% \\
$B^{+} \to \phi K^{*+} (K^{*+}\to K^{+}\pi^{0})$ 	 & 52       & $25.9^{+7.0}_{-6.3}$ & $50^{+13}_{-12}$\% \\
\hline 
\hline
\end{tabular}
\end{table}

\begin{figure}[!htb]
\centerline{
\resizebox*{2.0in}{1.692in}{
\includegraphics{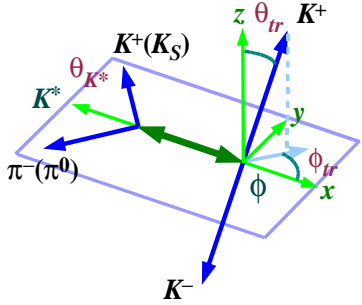}} }   
\caption{The definition of decay angles in $B \to \phi K^{*}$ decays.}
\label{fig:angular-def}
\end{figure}

The decay angles of a $B$-meson, to the two vector mesons $\phi$ and $K^{*}$ defined
in the transversity basis \cite{ref:transversity}, are shown in Fig.~\ref{fig:angular-def}.
The $x$-$y$ plane is defined to be the decay plane of $K^{*}$ and the $x$ axis is in
the direction of the $\phi$-meson.
The $y$ axis is perpendicular to the $x$ axis and is on the same side as the kaon from the $K^{*}$ decay.
The $z$ axis is perpendicular to the $x$-$y$ plane according to the right-hand rule,
$\theta_{\rm tr}$ ($\phi_{\rm tr}$) is the polar (azimuthal) angle 
with respect to the $z$-axis of the $K^+$ from $\phi$ decay 
in the $\phi$ rest frame, and $\theta_{K^*}$ is defined above.

The distribution of decays in the three angles \cite{ref:polarization_hepex}, $\theta_{K^*}$, $\theta_{\rm tr}$, and
$\phi_{\rm tr}$ is
\begin{eqnarray}
\label{equ:angularpdf}
\nonumber
&&{d^3 \Gamma (\phi_{\rm tr}, \cos\theta_{\rm tr}, \cos\theta_{K^*}) \over d\phi_{\rm tr} d\cos{\theta_{\rm tr}} d\cos{\theta_{K^*}}} 
= {9\over 32\pi} [ 
|A_\perp|^2 2 \cos^2{\theta_{\rm tr}} \sin^2{\theta_{K^*}} \\
\nonumber
&&~~~~~~ +|A_\parallel|^2 2 \sin^2{\theta_{\rm tr}} \sin^2{\phi_{\rm tr}} \sin^2{\theta_{K^*}} \\
\nonumber
&&~~~~~~ + |A_0|^2 4 \sin^2\theta_{\rm tr} \cos^2\phi_{\rm tr} \cos^2\theta_{K^*} \\
\nonumber
&&~~~~~~ +\sqrt{2} {\rm Re}(A^*_\parallel A_0) \sin^2\theta_{\rm tr}\sin 2 \phi_{\rm tr} \sin 2\theta_{K^*} \\
\nonumber
&&~~~~~~ -\eta\sqrt{2} {\rm Im}(A_0^* A_\perp) \sin 2\theta_{\rm tr} \cos\phi_{\rm tr} \sin 2\theta_{K^*} \\
&&~~~~~~ -2\eta{\rm Im}(A_\parallel^*A_\perp) \sin 2 \theta_{\rm tr} \sin \phi_{\rm tr} \sin^2 \theta_{K^*} ]~,
\end{eqnarray}
where $A_0$, $A_\parallel$, and $A_\perp$ are the complex amplitudes of the
three helicity states in the transversity basis with the normalization condition
$|A_0|^2 + |A_\parallel|^2 + |A_\perp|^2 = 1$, and $\eta\equiv +1$ ($-1$) corresponds to
$B$ ($\overline{B}$) mesons. $A_0$ denotes the longitudinal polarization of the $\phi\to K^+K^-$
system and $A_\perp$ ($A_\parallel$) is the transverse polarization
along the $z$-axis ($y$-axis).
The value of $|A_\perp|^2$ ($|A_0|^2 + |A_\parallel|^2 \equiv 1-|A_\perp|^2$)
is the $CP$-odd ($CP$-even)
fraction in the decay $B \to \phi K^*$ \cite{ref:polarization_hepex}. 
The imaginary phases of the amplitudes are sensitive to
final state interactions (FSI). 
The presence of FSI results in phases that differ from either $0$ or $\pm\pi$.


\begin{figure}[!htb]
\centerline{
\resizebox*{3.5in}{1.4in}{
\includegraphics{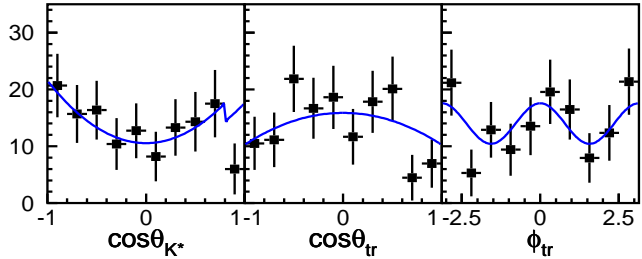}}
}   
\caption{ Projections of the transversity angles with results of the fit superimposed.
The points with error bars show the 
efficiency corrected data after subtraction of continuum and $B \to K^+K^-K^*$ background.
The discontinuity in the fit to $\cos\theta_{K^*}$ is due to the requirement of $\cos\theta_{K^*} < 0.8$
in $B^+ \to\phi K^{*+}(K^{*+}\to K^+\pi^0)$ decay.}
\label{fig:angular-proj}
\end{figure}

The complex amplitudes are determined by performing
an unbinned maximum likelihood fit \cite{ref:j/psiK*polar} 
with the $B \to \phi K^{*}$ candidates in the signal window.
The combined likelihood is given by
\begin{eqnarray}
\nonumber
&& \mathcal{L} =
\prod_i^N \epsilon(\theta_{K^*}, \theta_{\rm tr},\phi_{\rm tr}) [ f_{\phi K^{*}}
\cdot\Gamma(\theta_{K^*}, \theta_{\rm tr},\phi_{\rm tr}) \\
\nonumber
&& ~~~~~~~~
+ f_{q\overline{q}}\cdot R_{q\overline{q}}(\theta_{K^*}, \theta_{\rm tr},\phi_{\rm tr})\\
&& ~~~~~~~~
+ f_{KKK^{*}}\cdot R_{KKK^{*}}(\theta_{K^*}, \theta_{\rm tr},\phi_{\rm tr}) ]~ ,
\end{eqnarray}
where $\Gamma$ is the angular distribution function (ADF) given by Eq.~\ref{equ:angularpdf}, and
$R_{q\overline{q}}$ and $R_{KKK^{*}}$ are the
ADFs for continuum and $B \to KKK^{*}$ background, respectively.
The value of $\eta$ is determined from the charge of the kaon in the $K^{*}$ decay,
$R_{q\overline{q}}$ is determined from sideband data and 
$R_{KKK^{*}}$ is assumed to be flat. 
The detection efficiency function ($\epsilon$) is determined by MC.
The fractions of $\phi K^{*}$ ($f_{\phi K^{*}}$), $q\overline{q}$ ($f_{q\overline{q}}$) and
$B \to KKK^{*}$ ($f_{KKK^{*}}$) are parameterized as a function of $\Delta E$ and $M_{bc}$.
The value of $\arg(A_0)$ is set to zero and 
$|A_\parallel|^2$ is calculated from the normalization constraint in the fit.
The four parameters ($|A_0|^2$, $|A_\perp|^2$, $\arg(A_\parallel)$, and $\arg(A_\perp)$)
are determined from the fit.
The results from $B^0 \to \phi K^{*0}$ and $B^+ \to \phi K^{*+}$ may be combined if both
decays are dominated by $b \to s$ penguin transitions and
the $b \to u$ annihilation contribution to $B^+ \to \phi K^{*+}$ can be neglected.
Figure~\ref{fig:angular-proj} shows the angle distributions with projections
of the fit superimposed.
The amplitudes obtained from the fit are shown in Table~\ref{tab:amplitudes}.

The systematic uncertainties 
include the  slow pion detection efficiency ($6-7$\%), the background
from higher $K^*$ states ($2-8$\%), and the $B\to f_0 K^*$ background ($2-3$\%). 
The systematic uncertainty in the angular resolution is
estimated by MC simulation and found to be less than 1\%.
Uncertainties in the background PDFs, the signal yields, and the modeling of 
efficiency function ($\epsilon$) are estimated to be $1-3$\%.


\begin{table}[!htb]
\caption{The amplitudes obtained from the angular analysis, 
where the first errors are statistical and the second errors are systematic.}
\label{tab:amplitudes}
\begin{center}
\begin{tabular}{c|ccc}
\hline 
\hline 
Mode & $\phi K^{*0}$ & $\phi K^{*+}$ &  Combined \\
\hline
$|A_0|^2$	         & $0.52 \pm 0.07 \pm 0.05$  & $0.49 \pm 0.13 \pm 0.05$         & $0.51 \pm 0.06 \pm 0.04$  \\
$|A_\perp|^2$ 	         & $0.30 \pm 0.07 \pm 0.03$  & $0.12 ^{+0.11}_{-0.08} \pm 0.03$ & $0.24 \pm 0.06 \pm 0.03$  \\
$arg(A_\parallel)$ (rad) & $-2.30 \pm 0.28 \pm 0.04$ & $-2.07 \pm 0.34 \pm 0.07$        & $-2.21 \pm 0.22 \pm 0.05$ \\
$arg(A_\perp)$ (rad)     & $0.64 \pm 0.26 \pm 0.05$  & $0.93 ^{+0.55}_{-0.39} \pm 0.12$ & $0.72 \pm 0.21 \pm 0.06$  \\
\hline
\hline
\end{tabular}
\end{center}
\end{table}


The triple-product for a $B$ meson decay to two vector mesons takes the form 
$\overrightarrow{q} \cdot (\overrightarrow{\epsilon_1} \times \overrightarrow{\epsilon_2})$,
where $\overrightarrow{q}$ is the momentum of one of the vector mesons.
The other two vectors $\overrightarrow{\epsilon_1}$ and $\overrightarrow{\epsilon_2}$ are 
the polarizations of the two vector mesons. 
The presence of the triple product can be demonstrated by measuring a non-zero value of the asymmetry 
$A_T = {{\Gamma[\overrightarrow{v_1} \cdot (\overrightarrow{v_2} \times \overrightarrow{v_3})>0]
-\Gamma[\overrightarrow{v_1} \cdot (\overrightarrow{v_2} \times \overrightarrow{v_3})<0]} / {
\Gamma[\overrightarrow{v_1} \cdot (\overrightarrow{v_2} \times \overrightarrow{v_3})>0]+
\Gamma[\overrightarrow{v_1} \cdot (\overrightarrow{v_2} \times \overrightarrow{v_3})<0]}}$.
In the SM, $A_T$'s are very small. Experimentally, the
following two $T$-odd \cite{ref:TP-datta,ref:at12} quantities are used instead of $A_T$,
\begin{equation}
A_T^{0} \equiv {{\rm Im}(A_\perp A^*_0) \over A_0^2 + A_\perp^2 + A_\parallel^2}~,
~~~~
A_T^{\parallel} \equiv {{\rm Im}(A_\perp A^*_\parallel) \over A_0^2 + A_\perp^2 + A_\parallel^2}~.
\end{equation}
The comparison of these triple product asymmetries ($A_T^{0}$ and $A_T^{\parallel}$)
with the corresponding quantities for the $CP$-conjugated decays 
($\overline{A}_T^{0}$ and $\overline{A}_T^{\parallel}$) provide a 
true $T$-violating measurement. 

Additional variables that can be measured through angular analysis 
are suggested in Ref.~\cite{ref:NP-bounds}.
By introducing a time-dependent decay rate that can be written as follows
\begin{equation}
\Gamma(B(\overline{B}) \to V_1 V_2) = e^{-\Gamma t} \sum_{\lambda \le \sigma} 
\left( \Lambda_{\lambda\sigma} \pm \Sigma_{\lambda\sigma} \cos(\Delta M t) 
\mp \rho_{\lambda\sigma} \sin(\Delta Mt) \right) g_\lambda g_\sigma~,
\end{equation}
where $\Sigma$, $\Lambda$, and $\rho$ are expressed as
\begin{eqnarray}
\Lambda_{\lambda\lambda} = {1\over 2} ( |A_\lambda|^2 + |\overline{A}_\lambda|^2  )~, & &
\Sigma_{\lambda\lambda} = {1\over 2} ( |A_\lambda|^2 - |\overline{A}_\lambda|^2  )~,\\
\Lambda_{\perp i} = -{\rm Im} ( A_\perp A_i^* - \overline{A}_\perp \overline{A}_i^* )~, & &
\Sigma_{\perp i} = -{\rm Im} ( A_\perp A_i^* + \overline{A}_\perp \overline{A}_i^* )~,\\
\Lambda_{\parallel 0} = {\rm Re} ( A_\parallel A_0^* + \overline{A}_\parallel \overline{A}_0^* )~, & &
\Sigma_{\parallel 0} = {\rm Re} ( A_\parallel A_0^* - \overline{A}_\parallel \overline{A}_0^* )~,\\
\rho_{\perp\perp} = {\rm Im} \left({q\over p} A_\perp^* \overline{A}_\perp\right)~, & &
\rho_{ii} = -{\rm Im} \left({q\over p} A_i^* \overline{A}_i\right)~,\\
\rho_{\perp i} = {\rm Re}\left( {q\over p} [A_\perp^* \overline{A}_i + A_i^* \overline{A}_\perp] \right)~, & &
\rho_{\parallel 0} = -{\rm Im}\left( {q\over p} [A_\parallel^* \overline{A}_0 + A_0^* \overline{A}_\parallel] \right)~.
\end{eqnarray}
The subscript $i$ is one of the \{0, $\parallel$\} amplitudes,
$\Lambda_{\perp 0}$ and $\Lambda_{\perp\parallel}$ are equal to 
the T-violation parameters
$\mathcal{A}_T^{0} = \overline{A}_T^{0} - A_T^{0}$ and 
$\mathcal{A}_T^{\parallel} = \overline{A}_T^{\parallel} - A_T^{\parallel}$, respectively.
Since the parameters $\rho_{\lambda\sigma}$ appear in terms that are proportional to $\sin \Delta M t$, which 
vanish in a time integrated measurement, we only report results for the other twelve $\Lambda$ and $\Sigma$ parameters. 
The following equations should hold 
in the absence of NP:
\begin{eqnarray}
\Sigma_{\lambda\lambda} = 0~, &
\Sigma_{\parallel 0} = 0~, &
\Lambda_{\perp i} = 0~.
\end{eqnarray}
Any violation of these relations would be evidence for NP.

By separating $B$-meson and $\overline{B}$-meson samples and a rearranging the fitting parameters 
in the unbinned maximum likelihood fit, we obtain the helicity amplitudes for the $B$ and $\overline{B}$ samples,
the triple-product correlations, and other NP-sensitive  observables, which are given in
Table~\ref{tab:amplitudes-sep} and \ref{tab:NP-observables}.  


\begin{table}[!htb]
\caption{The measured helicity amplitudes and triple-product correlations
in the $B$-meson and $\overline{B}$-meson samples.}
\label{tab:amplitudes-sep}
\begin{center}
\begin{tabular}{c|cc}
\hline 
\hline 
Mode & $B$-meson combined ($\phi K^{*0}$ and $\phi K^{*+}$) & $\overline{B}$-meson combined ($\phi \overline{K}^{*0}$ and $\phi K^{*-}$) \\
\hline
$|A_0|^2$	         & $0.41 \pm 0.10 \pm 0.03$  & $0.59 \pm 0.08 \pm 0.06$ \\
$|A_\perp|^2$ 	         & $0.24 \pm 0.10 \pm 0.02$  & $0.26 \pm 0.08 \pm 0.04$ \\
$arg(A_\parallel)$ (rad) & $-2.29 \pm 0.35 \pm 0.13$ & $-2.05 \pm 0.31 \pm 0.04$ \\
$arg(A_\perp)$ (rad) 	 & $0.74 \pm 0.31 \pm 0.10$  & $0.81 \pm 0.31 \pm 0.06$ \\
\hline
$A_T^{0}$         & $0.21\pm0.08\pm0.03$  & $0.28\pm0.09\pm0.05$ \\
$A_T^{\parallel}$ & $0.04\pm0.08\pm0.02$  & $0.06\pm0.06\pm0.02$ \\
\hline
\hline
\end{tabular}
\end{center}
\end{table}


\begin{table}[!htb]
\caption{The observables sensitive to NP extracted from angular analysis.}
\label{tab:NP-observables}
\begin{center}
\begin{tabular}{ccr|ccr}
\hline 
\hline 
\multicolumn{3}{c|}{$\Lambda$ Observables} & \multicolumn{3}{c}{$\Sigma$ Observables} \\ 
\hline
~~$\Lambda_{00}$					    &=& $ 0.50\pm0.06\pm0.04$ ~~&~~ $\Sigma_{00}$		  &=& $-0.09\pm0.06\pm0.02$ ~~\\
~~$\Lambda_{\parallel\parallel}$			    &=& $ 0.25\pm0.06\pm0.02$ ~~&~~ $\Sigma_{\parallel\parallel}$ &=& $ 0.10\pm0.06\pm0.02$ ~~\\
~~$\Lambda_{\perp\perp}$				    &=& $ 0.25\pm0.06\pm0.03$ ~~&~~ $\Sigma_{\perp\perp}$	  &=& $-0.01\pm0.06\pm0.02$ ~~\\
~~$\Lambda_{\perp 0}$ ($=\mathcal{A}_T^{0}$)		    &=& $ 0.07\pm0.11\pm0.04$ ~~&~~ $\Sigma_{\perp 0}$       	  &=& $-0.49\pm0.12\pm0.07$ ~~\\
~~$\Lambda_{\perp\parallel}$ ($=\mathcal{A}_T^{\parallel}$) &=& $ 0.02\pm0.10\pm0.03$ ~~&~~ $\Sigma_{\perp\parallel}$	  &=& $-0.09\pm0.10\pm0.02$ ~~\\
~~$\Lambda_{\parallel 0}$				    &=& $-0.39\pm0.13\pm0.06$ ~~&~~ $\Sigma_{\parallel 0}$	  &=& $-0.11\pm0.13\pm0.04$ ~~\\
\hline
\hline
\end{tabular}
\end{center}
\end{table}

In summary,
the decay amplitudes for $B \to \phi K^{*}$ are determined 
by fitting the angular distributions in the transversity basis.
The longitudinal polarization component ($|A_0|^2$) is in agreement with Ref.~\cite{ref:phik_prl}, but differs
from the naive SM prediction.
The measured value of $|A_\perp|^2$ 
shows that both $CP$-odd ($|A_\perp|^2$) and $CP$-even ($|A_0|^2 + |A_\parallel|^2$) 
components are present in $\phi K^*$ decays with a ratio of 1:3.
The phase of $A_\perp$ and $A_\parallel$ differ from zero or $-\pi$
by 3.6 and 3.7 standard deviations \cite{ref:statistical-only}, respectively. Thus, our data indicate
the presence of final state interactions.
The triple products $A_T^{0}$ and $\overline{A}_T^{0}$ differ from zero by  
2.4$\sigma$ and 2.8$\sigma$, respectively \cite{ref:statistical-only}. This may 
also be an indication of the presence of final state interactions. However, the differences 
between $A_T$'s and $\overline{A}_T$'s (= $\mathcal{A}_T^{0,\parallel}$ = $\Lambda_{\perp 0,\parallel}$)
that are sensitive to $T$-violating asymmetry
are consistent with zero as well as \cite{ref:babar_ex0406063}.
The equations, $\Sigma_{\lambda\lambda} = 0$, $\Sigma_{\parallel 0} = 0$, and
$\Lambda_{\perp i} = 0$, should hold 
in the absence of NP. Our data does not show any significant violation of those
relations.


We thank the KEKB group for the excellent operation of the
accelerator, the KEK Cryogenics group for the efficient
operation of the solenoid, and the KEK computer group and
the National Institute of Informatics for valuable computing
and Super-SINET network support. We acknowledge support from
the Ministry of Education, Culture, Sports, Science, and
Technology of Japan and the Japan Society for the Promotion
of Science; the Australian Research Council and the
Australian Department of Education, Science and Training;
the National Science Foundation of China under contract
No.~10175071; the Department of Science and Technology of
India; the BK21 program of the Ministry of Education of
Korea and the CHEP SRC program of the Korea Science and
Engineering Foundation; the Polish State Committee for
Scientific Research under contract No.~2P03B 01324; the
Ministry of Science and Technology of the Russian
Federation; the Ministry of Education, Science and Sport of
the Republic of Slovenia; the National Science Council and
the Ministry of Education of Taiwan; and the U.S.\
Department of Energy.



\begin{thebibliography}{99}

\bibitem{ref:ckm}
{N. Cabibbo, Phys. Rev. Lett. {\bf 10}, 531 (1963);
 M. Kobayashi and T. Maskawa, Prog. Theor. Phys. {\bf 49}, 652 (1973).}

\bibitem{ref:TP-datta}
{A. Datta and D. London, hep-ph/0303159.}

\bibitem{ref:NP-bounds}
{D.~London, N.~Sinha and R.~Sinha. hep-ph/0402214.}

\bibitem{ref:phik_prl}
{Belle Collaboration, K.-F.Chen, {\it et al.}, Phys. Rev. Lett. {\bf 91}, 201801 (2003)} 
      
\bibitem{ref:babar_ex0406063}
{{\sc BaBar} Collaboration, J.G. Smith, hep-ex/0406063.}





\bibitem{ref:Belle}
{A. Abashian {\it et al.}, Nucl. Instr. Meth. A479, 117 (2002).}

\bibitem{ref:KEKB}
{S. Kurokawa, E. Kikutani, Nucl. Instr. Meth. A499, 1 (2003).}
   
\bibitem{ref:continuum_suppression}
{We use the $S_{\perp}$ variable as defined in
CLEO Collaboration, R. Ammar {\it et al.}, Phys. Rev. Lett. {\bf 71}, 674 (1993),
and the thrust angle and modified Fox-Wolfram
moments defined in Belle Collaboration, K. Abe {\it et al.}, Phys. Lett. B {\bf 517}, 309 (2001).}   

\bibitem{ref:tagging}
H. Kakuno {\it et al.}, hep-ex/0403022.

\bibitem{ref:range}
{The range corresponds to the range of values for different
sub-modes. This convention is used throughout this letter.}

\bibitem{ref:argus}
{The functional form is $x \sqrt{1-x^{2}}\exp(\alpha(1-x^{2}))$, where 
$x = {M_{bc}}/{E_{\rm beam}}$. 
ARGUS Collaboration, H. Albrecht {\it et al.} , Phys. Lett. B {\bf 241} (1990) 278; {\bf 254} (1991) 288.} 

\bibitem{ref:PDG}
{Particle Data Group, K. Hagiwara {\it et al.}, Phys. Rev. D {\bf 66} (2002).}

\bibitem{ref:transversity}
{I. Dunietz, H. Quinn, A. Snyder, W. Toki, and H.J. Lipkin, Phys. Rev. D {\bf 43}, 2193 (1991).}

\bibitem{ref:polarization_hepex}
{K. Abe, M. Satpathy and H. Yamamoto, hep-ex/0103002 (2001).}

\bibitem{ref:j/psiK*polar}
{Belle Collaboration, K. Abe {\it et al.}, Phys. Lett. B {\bf 538}, 11 (2002)}

\bibitem{ref:at12}
{We take the definitions of $A_T^0 = A_T^{(1)}$ and 
$A_T^\parallel = A_T^{(2)}$, but $\overline{A}_T^0 = -\overline{A}_T^{(1)}$ and 
$\overline{A}_T^\parallel = -\overline{A}_T^{(2)}$. The variables $A_T^{(1)}$ and $A_T^{(2)}$
are defined in Ref.~\cite{ref:TP-datta}.}





\bibitem{ref:statistical-only}
{The significance here is defined as $\sqrt{-2\ln(\mathcal{L}/\mathcal{L}_{max})}$. The effects
of systematic uncertainties are not taken into account.}
     
\end{thebibliography}
\end{document}